**Nature and Origin of Unusual Properties in Chemically Exfoliated 2D MoS$_2$**

Debasmita Pariari and D. D. Sarma [a]

*Solid State and Structural Chemistry Unit, Indian Institute of Science, Bengaluru – 560012*

[a] Corresponding author: sarma@iisc.ac.in

**Abstract:**

MoS$_2$ in its two-dimensional (2D) form is known to exhibit many fundamentally interesting and technologically important properties. One of the most popular routes to form extensive amount of such 2D samples is the chemical exfoliation route. However, the nature and origin of the specific polymorph of MoS$_2$ primarily responsible for such spectacular properties has remained controversial with claims of both T and T′ phases as well as metallic and semiconducting natures. We show that a comprehensive scrutiny of the available literature data of Raman spectra from such samples allow little scope for such ambiguities, providing overwhelming evidence for the formation of the T′ phase as the dominant metastable state in all such samples. We also explain that this small band-gap T′ phase may attain substantial conductivity due to thermal and chemical doping of charge-carriers, explaining the contradictory claims of metallic and semiconducting nature of such samples, thereby attaining a consistent view of all reports available so far.

**Introduction:**

The discovery of atomically thin layer of graphene from three dimensional graphite crystal by Geim and Novoselov in 2004,[1] opened up a new avenue of research in two dimensional (2D) layered materials. Overwhelming attention has been focused on the study of the analogous layered materials since then. Among these the long known, well-studied and technologically important is molybdenum disulphide ($MoS_2$), a member of the transition metal dichalcogenides (TMDs) family. Although investigations on $MoS_2$ can be traced through decades due to its natural abundance in the earth's crust, important catalytic properties[2–4] and extensive usage as solid state lubricant[5–7], it has seen an exponential increase in number of publications recently.[8] Unlike graphene, each layer of $MoS_2$ is three atomic layer thick with a thickness of 6.2 Å,[9] in which the planes of Mo atoms are sandwiched between two atomic layers of S with strong in-plane covalent bonding and between Mo and S planes, while such layers of $MoS_2$ with three atomic planes are vertically stacked *via* weak van der Waals interactions. This allows for easy mechanical exfoliation of single or few layers of $MoS_2$, ideal for investigating 2D form of $MoS_2$. The large van der Waals gap can also allow different ions to readily intercalate between the $MoS_2$ layers.[8] This route has been often used to chemically exfoliate $MoS_2$ into the 2D form, since such intercalation typically expands the interlayer separation greatly, reducing the coupling between successive $MoS_2$ layers to an insignificant level.

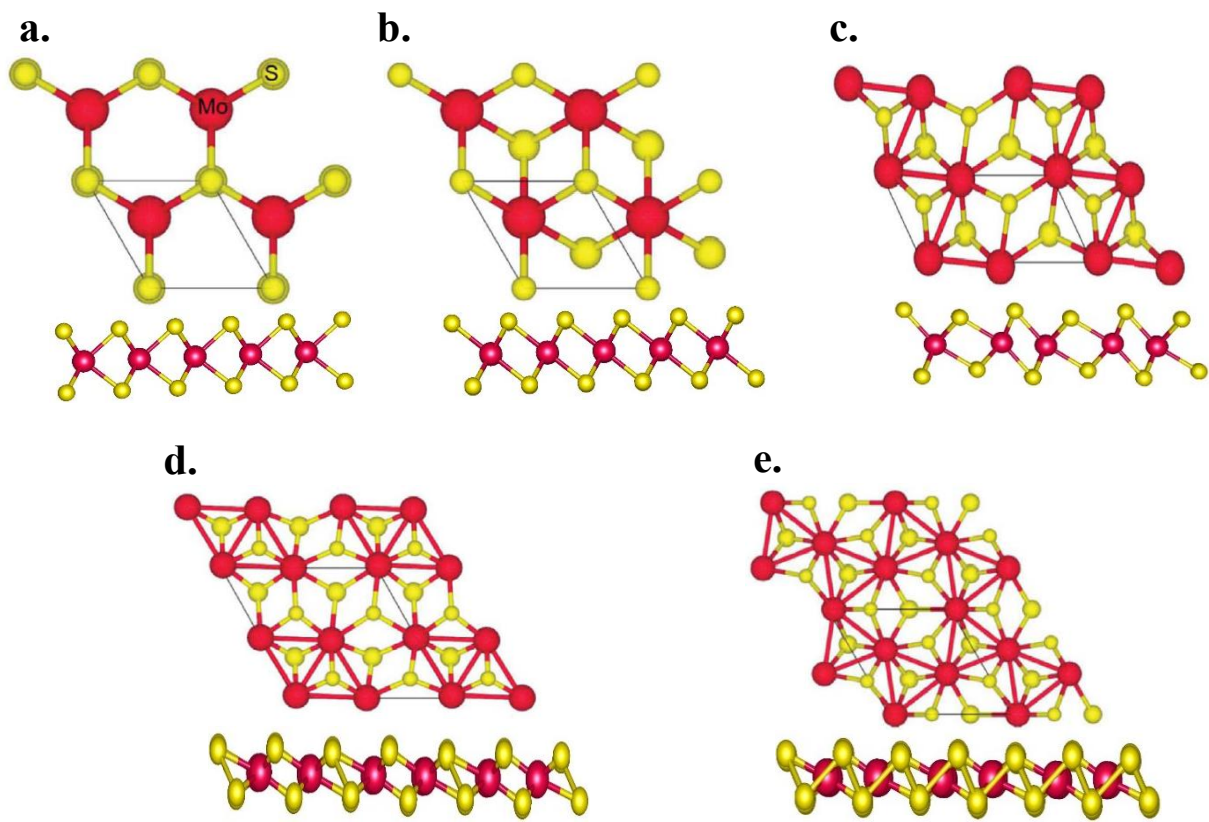

**FIG. 1.** Top and side view of **(a)** H, **(b)** T, **(c)** (1 × 2) supershell, T′, **(d)** (2 × 2) supershell, T″ and **(e)** (√3 × √3) supershell, T‴ phases of $MoS_2$. The short Mo-Mo bonds are shown in red for panels (c)-(e). Reproduced from ref. 32, © 2017 APS.

Bulk MoS$_2$ is semiconducting with an indirect band-gap of 1.2 eV[10] whereas a monolayer of MoS$_2$ is a direct band-gap (1.8 eV) semiconductor.[11] One of the interesting features of MoS$_2$ is that it can exist in several polymorphic forms, shown in Fig. 1, depending on how the three hexagonal layers of S-Mo-S are stacked above each other. A-B-A type of stacking, with the top and the bottom S layers being directly above each other, gives rise to the thermodynamically stable polymorph H (Fig. 1a), with six S atoms oriented around the central Mo atom in a trigonal prismatic coordination. In contrast, the unstable T form, shown in Fig. 1b, has the A-B-C type stacking with an octahedral coordination of S atoms around Mo.[8] This T polymorph can undergo various Jahn-Teller type distortions leading to the formation of superlattices with different metal-metal clustering patterns, such as $a_0 \times 2a_0$ with dimerized zig-zag Mo chains (T′) in Fig. 1c, $2a_0 \times 2a_0$ with tetramer Mo-Mo clusters in a diamond formation (T″) in Fig. 1d and $\sqrt{3}a_0 \times \sqrt{3}a_0$ with a trimerized clustering (T‴) in Fig. 1e.[12,13] Such diverse polymorphic forms are of great importance, since their electronic properties vary greatly, with the metastable T′, T″ and T‴ phases being semiconductors with varying band-gaps and the T phase being metallic. Because of this tuneability of electronic properties, ranging from wide gap insulator to metal, MoS$_2$ has emerged as a potential candidate for an extraordinarily diverse range of novel applications in different fields, such as transistors,[14,15] optoelectronics,[16] catalysis,[2,17–19] photodetectors,[20] supercapacitors,[21] secondary batteries,[22,23] and even as superconductors.[24,25]

MoS$_2$ can be easily transformed to its various metastable states using different routes. These have been extensively studied and reported in literature, such as, plasma hot electron transfer,[26,27] mechanical strain,[28,29] and electron-beam irradiation.[30,31] However, the chemical routes to achieve such transformation have proven to be the most facile and, therefore, popular ones. Chemical routes in turn involve chemical,[32,33] electrochemical alkali metal intercalation,[34,35] or expansion of the interlayer distance by hydrothermal synthesis.[36,37] Although, through all the above-mentioned processes, the stable H phase is known to be transformed to one of the metastable states, the structure and electronic properties of the resultant phase have still remained highly contentious with many conflicting claims and ambiguities. Theoretical calculations predict that the Jahn-Teller distorted T′ and T″ are small band-gap semiconductors[38,39] and T‴ is a ferroelectric insulator.[40] Interestingly, the undistorted T phase is theoretically predicted to be dynamically unstable as phonon dispersion of this phase shows an instability at the zone boundary.[38] Despite such distinct properties expected of each variant, most of the experimental papers, dealing with such chemically treated samples do not clearly identify the specific phase formed, often using the term T or in few cases T′ in a generic manner to denote a metastable phase. There are also several reports where instead of identifying any crystallographic phase, the additional phases formed due to such chemical treatments are classified by their presumed electronic or transport properties and termed as metallic or semiconducting MoS$_2$.[33,41–43] Unfortunately, the generic use of T to denote the metastable form and the frequent claim of a metallic nature have created an impression in the community that the metastable state formed is predominantly the metallic, undistorted

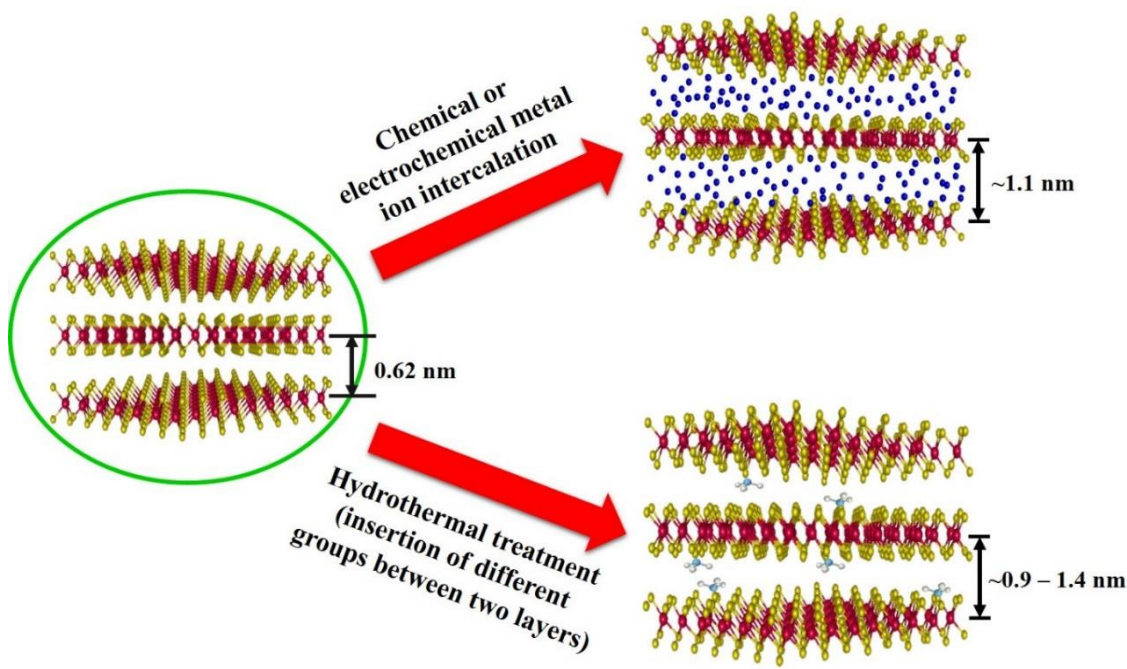

**FIG. 2.** Schematics of different chemical exfoliation techniques.

T phase and not one of the small band-gap semiconducting, distorted T′, T″ or T‴ phase. We critically scrutinize this dogma by looking at all relevant data already available in literature to arrive at the contrarian view in this perspective.

**Chemical exfoliation of MoS$_2$:**

Briefly, there are three distinct chemical exfoliation routes employed for MoS$_2$, namely chemical intercalation, electrochemical intercalation and hydrothermal or closely related solvothermal synthesis. Schematic representations of these two routes are shown in Fig. 2. Historically, intercalation has been applied to layered materials as a means of exfoliating individual 2D layers from their bulk counterparts in large quantities. Intercalation chemistry plays a key role in a majority of the liquid-based exfoliation methods which in contrast to the mechanical exfoliation, presents great advantages for the mass production of 2D materials.[44,45] The key principle for the intercalation-based exfoliation is to increase the interlayer spacing between individual layers by inserting foreign species. This weakens further the already weak interlayer van der Waals interaction and reduce the energy barrier of exfoliation.[44]

Although some research on the intercalation of different alkali metals into MoS$_2$ has been reported,[46–49] most of the attention has been focused on the intercalation of lithium (Li). This is based on the expectation that Li$^+$ ions, with the smallest ionic radius among all alkali metal ions, will easily enter the interlayer space and also because of the potential of such Li-intercalated materials as components of high-power rechargeable batteries. The chemical route of Li-intercalation, developed by Joensen and co-workers, involves treating MoS$_2$ with *n*-butyl lithium (*n*-BuLi) in hexane as the intercalating agent followed by a water exfoliation step.[50] Schematic illustration of the procedure is shown in Fig. 3a. The key step of this procedure is the formation of Li$_x$MoS$_2$ *via* a slow process, requiring Li-intercalation about 48 hours or more. The lithiated solid product is retrieved by filtration and washed with hexane to remove excess Li and organic residues of *n*-BuLi. In the next step of washing with water, the bare alkali ions are immediately solvated by water molecules which form number of layers in the Van der Waals gap facilitating the exfoliation process and also stabilizing the monolayers in the solvent.

However, for the solvated phase, the Li content is significantly lower than in the intercalated compounds prior to washing with water.[51] For the solvated phases of these type of compounds, the alkali metals remain almost fully ionised and the guest (alkali atoms) and host charges (residual negative charges on disulphide layers) remain separated by solvent layers. In the solvated phase, the expansion of interlayer spacing with respect to that of the pristine compound naturally depends on the number of solvent layers formed in the interlayer space, which in turn depends on the ionic radius of the intercalated alkali metal.

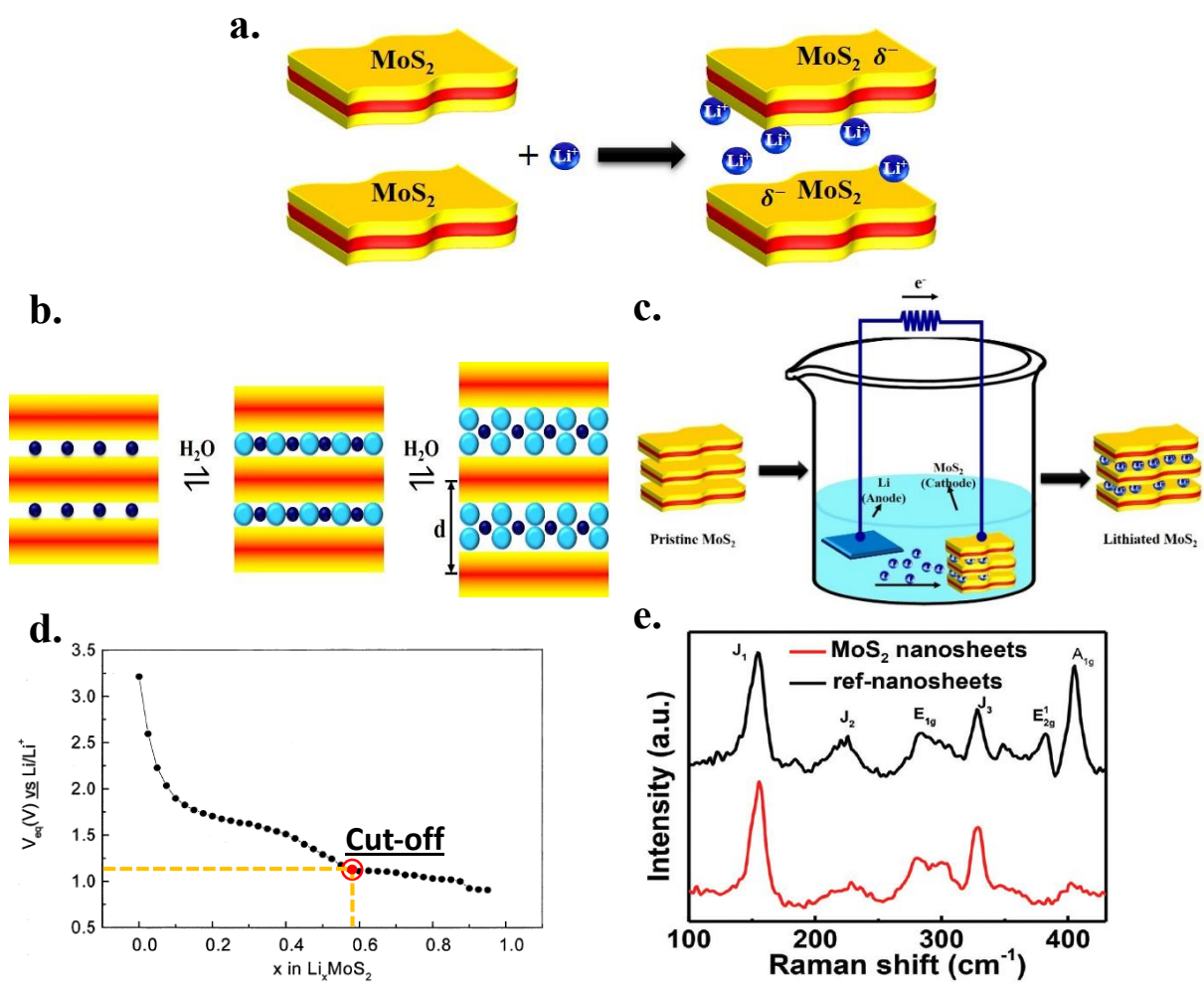

**FIG. 3.** **(a)** Schematics of lithium ion intercalation into the MoS$_2$ matrix; **(b)** Representation of the intercalated alkali metal cation with water molecules arranged in monolayer and bilayer; **(c)** Schematics of the electrochemical intercalation method; **(d)** Voltage-composition curve for the intercalation of lithium into MoS$_2$ to form Li$_x$MoS$_2$. Cut-off voltage is shown with red circle beyond which Li$_x$MoS$_2$ decomposes forming Li$_2$S; **(e)** Pure T' phase of MoS$_2$ prepared through the formation of Li$_x$MoS$_2$ via high temperature solid state synthesis followed by water exfoliation. Figures reproduced with permission from: **(d)**, ref. 56, © 2002 Elsevier; **(e)**, ref. 74, © 2017 RSC.

Thus, the distance between the adjacent layers change in a stepwise manner, depending on whether the solvating molecules form a monolayer or bilayer, as demonstrated in Fig. 3b.[51,52] The hydration energy, characterised by charge/radius ratio, of Li$^+$ and Na$^+$ is greater due to the smaller radius compared to the other alkali metal ions, leading to the formation of two water layers in the intercalation compound for these two guest ions, whereas all the other cations stabilize with mono-layered packing of water in the interlayer space. While we focus our discussion primarily on Li intercalation route in this article, in view of its pre-eminence in the published literature as the most preferred route, we note in passing many other investigations of chemical intercalation driven exfoliation of MoS$_2$, involving other alkali ions.[49,53,54]

Electrochemical intercalation allows a considerably higher control on the amount of Li-intercalated, while also achieving a faster rate of intercalation for small quantities of the host, compared to the chemical intercalation route. In general, the Li$^+$ electrochemical intercalation (see Fig. 3c) is performed in a test cell using a Li foil as the anode, LiPF$_6$ or LiClO$_4$ in propylene carbonate as the electrolyte, and MoS$_2$ as the cathode using galvanostatic discharge at a certain current density.[55] The advantage of this method is that Li$^+$ insertion can be monitored and precisely controlled, so that the galvanostatic discharge can be stopped at the desired Li content to avoid decomposition of the Li-intercalated compounds by optimizing the cut-off voltage. Fig. 3d shows a typical voltage–composition curve for the intercalation of Li in H-MoS$_2$ with the cut-off voltage shown with the red dot.[56]

Hydrothermal[57] and solvothermal[58,59] synthesis are two other, closely interrelated, popular methods for chemical exfoliation of MoS$_2$ by forcing small molecules into the interlayer gap leading to the weakening of the van der Waals interaction between two adjacent layers. Typically, in this method, molybdate is used to react with a sulphide or sulphur in a stainless-steel autoclave in presence of a reducing agent, leading to a series of physicochemical reactions under relatively high temperature (~ 200°C) and high pressure for several hours. In most of the cases, the resultant MoS$_2$ with greater interlayer spacing, is intercalated with NH$_4^{+}$[36,60] or H$_2$O[43] molecules.

**Vibrational structure of different phases:**

Raman spectroscopy has been used routinely in the literature to differentiate different polymorphs of MoS$_2$, as it provides a reliable methodology based on the distinctive vibrational structures expected from each phase. Any spectroscopic, including Raman, transition probability is proportional to the square of the corresponding transition matrix element, given by,

$$O_{fi} = \langle \psi_f | \hat{O} | \psi_i \rangle \qquad \ldots (1)$$

where, $\psi_i$ and $\psi_f$ are the wavefunctions of the initial and final states of the transition and $\hat{O}$ is the transition operator. For an allowed transition, the integral in equation (1) has to be non-zero, which can be simply translated in group theoretical terms by mentioning that for an allowed transition the direct product of the irreducible representation of the corresponding terms in equation (1) has to contain a function that forms a basis for the totally symmetric irreducible representation. In other words, $\Gamma_f \otimes \Gamma_O \otimes \Gamma_i$ must contain the totally symmetric irreducible representation for the transition matrix element to be nonzero. The vibrational ground state $\langle\psi_i|$ of a molecule, relevant for Raman transitions, always transforms as the totally symmetric irreducible representation; therefore, the above general condition, in the context of Raman spectroscopy, requires that the direct product of $\Gamma_f \otimes \Gamma_O$ has to contain a totally symmetric irreducible representation. For Raman Spectroscopy $\hat{O}$ is the polazibility operator. The irreducible representation ($\Gamma_{vib}$) corresponding to the normal modes of vibrations at the zone centre (k = 0) of the point group $D_{3h}$ describing the structure of monolayer $MoS_2$ in its H phase[61], is given by,

$$\Gamma_{vib}(H) = A'_1 + 2E' + 2A''_2 + E'' \qquad \ldots (2)$$

Careful inspection of the basis functions of each irreducible representation of $D_{3h}$ point group reveals that $A'_1$, $E''$ and one of $E'$ are the Raman active modes which give rise to three peaks in the Raman spectra of this phase.

By carrying out similar analysis on T ($D_{3d}$ point group) and T′ ($C_{2h}$ point group) phase, one immediately finds out that the irreducible representation of the normal modes of vibrations are,[61,62]

$$\Gamma_{vib}(T) = A_{1g} + E_g + 2E_u + 2A_{2u} \qquad \ldots (3)$$

$$\Gamma_{vib}(T') = 6A_g + 3B_g + 3A_u + 6B_u \qquad \ldots (4)$$

with the Raman active modes as $A_{1g}$, $E_g$ for the T phase (2 peaks) and $A_g$, $B_g$ (9 peaks) for the T′ phase. We note that this analysis does not provide any indication of how intense or weak a specific symmetry-allowed Raman signal may be; therefore, it is entirely possible that a symmetry-allowed Raman signal is not observed in an experiment due to its low intensity. In this sense, the above consideration helps to establish a rigorous upper limit on the number of peaks one may observe for a given phase of $MoS_2$ and not the lower limit. Before turning to the available information on Raman spectra of these samples in the literature, we note the reason to consider the undistorted T phase as a possible candidate, despite the undeniable theoretical result that this is an unstable phase that will spontaneously distort itself into one of the lower energy T′, T″ or T‴ phases; in other words, theoretical analysis shows that it is purely unstable and cannot be a metastable state. However, such theoretical analysis bases itself on the long-range periodic structure of the T phase, whereas the observed metastable states of $MoS_2$ coexist as small patches within the domains of the H phase; this leads to several additional effects, such as the finite size effect, strains generated across the two phase boundaries and possibilities of charge doping, making the real scenario very different from the idealized case considered by theoretical approaches; many of these additional effects may have the ability to make the finite-sized and embedded and/or charge-doped T phase a metastable rather than an unstable phase under the experimental realization.

**Table 1:** Theoretically calculated and experimentally obtained peak position (in cm$^{-1}$) in the Raman spectra of MoS$_2$ in its H phase.

| H | 1$^{st}$ ($E''/E_{1g}$) | 2$^{nd}$ ($E'/E^1_{2g}$) | 3$^{rd}$ ($A'_1/A_{1g}$) |
|---|---|---|---|
| **Theoretical calculation** | 280 - 286 [12,63] | 375 - 385 [12,38,63] | 402 – 408 [12,38,63] |
| **Experimental value** | 286 (very low intensity) [61,64] | 380 – 383 [61,64–67] | 402 – 409 [61,64–67] |

**Table 2:** Theoretically calculated and experimentally obtained peak position (in cm$^{-1}$) in the Raman spectra available in the literature for MoS$_2$ in its different polymorphic phases. The first group of data are from theoretical calculations for pure H, T and T' phases, as indicated, and serve as reference values. Subsequent entries are for experimentally obtained Raman peak positions along with the claimed phases in corresponding references with our assignments appearing in the last column under "Phases". In case of a mixture of phases, the dominant one appears first. The publication years are shown in bold within parentheses for the subsequent group of reports.

| Ref. | Peak positions (cm$^{-1}$) | | | | | | | | | | | Claimed as | Phase |
|---|---|---|---|---|---|---|---|---|---|---|---|---|---|
| 12,38,61,68,69 | | | | | | 280 - 286 | | | | 375 - 385 | 402 - 409 | H | |
| 12,61,69 | | | | | 258 - 268 | | | | 356 - 391 | | | T | |
| 12,38,61,69–71 | 138 - 140 | 146 - 160 | 200 - 209 | 216 - 230 | | 286 - 300 | 333 - 336 | 350 | 399 | | 412 | T' | |
| **(2011)** | | | | | | | | | | | | | |
| 33 | | 151 | | 229 | | 300 | 332 | | | 382 | 405 | T + H | H + T' |
| **(2013)** | | | | | | | | | | | | | |

| | | | | | | | | | | | | |
|---|---|---|---|---|---|---|---|---|---|---|---|---|
| 42 | 150 | | 219 | | | 327 | | | 382 | 406 | | T+H | H+T' |
| **(2014)** | | | | | | | | | | | | | |
| 72 | | 200 | 225 | | | | 353 | | 380 | 405 | | T | H+T' |
| **(2015)** | | | | | | | | | | | | | |
| 73 | 156 | | 226 | | 299 | 333 | | | | 405 | | T+H | T'+H |
| 60 | 156 | | 226 | | 284 | 333 | | | 377 | 407 | | T+H | T'+H |
| 34 | 150 | 200 | | | | | | | 380 | 408 | | T+H | T'+H |
| **(2016)** | | | | | | | | | | | | | |
| 43 | 146 | | 219 | | 283 | 326 | | | | 404 | | T+H Metallic | T'+H |
| 67 | 158 | | 218 | | | 334 | | | 383 | 409 | | T'+H | T'+H |
| **(2017)** | | | | | | | | | | | | | |
| 59 | 150 | | | | | 320 | | | 380 | 405 | | T+H | T'+H |
| 74 | 157 | | 229 | | 283 | 330 | | | | 403 | | T' | T'+H |
| 37 | | | 235 | | 280 | 336 | | | 375 | 404 | | T+H | T'+H |
| 25 | 156 | | 228 | | 283 | 330 | | | | 403 | | T'+H | T'+H |
| 48 | 154 | | 219 | | | 327 | | | 380 | 404 | | T+H | T'+H |
| 32 | 156 | | 227 | | | 330 | | | 383 | 405.4 | | T+H | T'+H |
| 36 | 153.2 | | 226.4 | | | 336.7 | | | 381.9 | 406.9 | | T+H | T'+H |

| | | | | | | | | | | | |
|---|---|---|---|---|---|---|---|---|---|---|---|
| **(2018)** | | | | | | | | | | | |
| 65 | | 156 | | 218 | 283 | 333 | | | | 408 | | T′+H | T′+H |
| 62 | | 156 | | 228 | | 330 | | | 383 | 409 | | T′+H | T′+H |
| **(2019)** | | | | | | | | | | | |
| 75 | | 187 | | 224 | 289 | | | | 380 | 405 | | T+T′ | T′+H |
| 76 | | 156 | | ~230 | ~290 | ~330 | | | | | | T′+H | T′+H |
| 77 | | 153.2 | | 226.5 | | 336.8 | | | ~380 | ~409 | | T+H | T′+H |
| **(2020)** | | | | | | | | | | | |
| 78 | | 156 | | 220 | 283 | 330 | | | 380 | 404 | | T′+H | T′+H |

The above discussion on the maximum number of Raman peaks expected in any given phase of MoS$_2$ is already a powerful tool to probe the possible phases formed. This is further aided by detailed quantum mechanical calculations of the phonon spectrum of each phase, providing quantitative estimates of the various peak positions. For example, in Table 1, we have tabulated the theoretically calculated frequencies for the three Raman active modes, namely, $E_{1g}$, $E_{2g}^1$, $A_{1g}$ in the H phase of MoS$_2$, from refs. 12, 38 and 63. We have also shown the experimentally obtained estimates of the Raman peak positions for the H phase. The remarkable agreement between the experiment and the calculated values provide us with confidence in determining the specific phases of MoS$_2$ present in any given sample of MoS$_2$ from a scrutiny of its Raman spectrum. With this aim in mind, we have collected every publication on chemically exfoliated MoS$_2$ that also reports the corresponding Raman spectrum. We summarize the comprehensive information available in the literature by tabulating the peak positions of Raman spectrum in each such publication together with the assignment of the nature of the metastable phase of MoS$_2$, suggested in that publication in Table 2. The top three rows of this table provide the summary of all theoretically calculated peak positions, reported for various polymorphs of MoS$_2$ so far in the literature and we use these values for our own phase identifications of the reported spectra in the last column of the table. When we write T′ + H in the last column, we imply that the most intense signal of the Raman spectrum reported in that reference arises from the T′ phase while there are also lower intensity signals present that are due to the presence of the H phase; H + T′ implies exactly the opposite scenario. We have arranged the references in the chronological order and grouped them under the year of publication. For a quicker comprehension of all these disparate data sets in Table 2, we have represented these results also in the form of a plot in Fig. 4. Table 2 and Fig. 4 together make evident a few interesting observations. First, it appears that the claim of the T phase formation was relatively more abundant till about 2017, while the claim of the formation of T′ phase has become relatively more frequent in recent years. We also note that most of the Raman spectra reported in the literature invariably exhibits signatures of the H phase with peaks appearing at ~382 and ~405 cm$^{-1}$. This suggests that the conversion of bulk MoS$_2$ to its few layered 2D form *via* chemical exfoliation generally does not lead to a complete transformation of the stable H phase to its various metastable T forms. Most importantly, we find that all Raman spectra published so far to provide evidence of metastable states in chemically exfoliated, few layer MoS$_2$ have three or more Raman peaks in addition to those attributable to the presence of H phase in the sample. One notable exception is achieved *via* a high temperature solid state synthesis of Li$_x$MoS$_2$ and successive exfoliation in aqueous acidic solution, leading to the formation of homogeneous T′ phase instead of the mixtures of H, T and T′ phases.[74] The Raman spectra of this sample is compared with that of a sample synthesized through the traditional intercalation route using n-BuLi in Fig. 3e. The complete absence of the $E^1_{2g}$ mode (383 cm$^{-1}$), characteristic of the H phase, establishes the absence of any H phase in this sample, while the signal due to the E$_{2g}$$^1$ mode is clearly visible in the *n*-BuLi treated sample mentioned as ref-nanosheets in Fig. 3e. The clear observation of more than two peaks from the metastable phase in every reported case of Raman spectra, as shown in Table 2, establishes the absence of any significant extent of the undistorted, metallic T phase formation in the chemical exfoliation route; this is not entirely surprising in view of the intrinsic instability of the T phase. What is surprising, however, is the often reported[79–81] metallic nature of the chemically exfoliated MoS$_2$ samples, since the H phase, as well as all metastable distorted phases, such as T′, T″ and T‴, are known to be semiconducting. Since the question of metallic/insulating behaviour of any sample is intrinsically connected with its electronic structure, next we turn to the discussion of electronic structures of such samples.

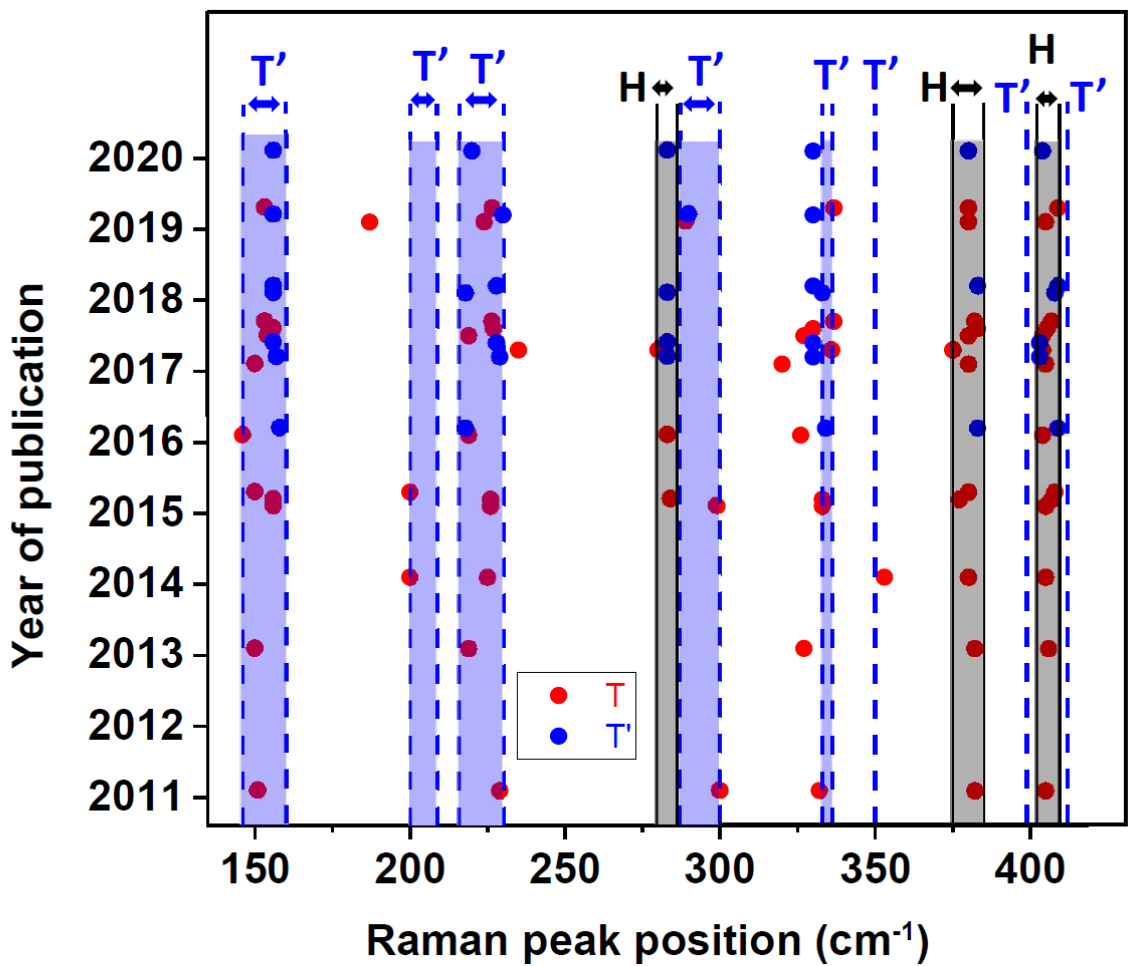

**FIG. 4.** A graphical representation of the results of all Raman studies available in the literature and tabulated in Table 2. The peak positions of all Raman signals reported in any given publication appears along a horizontal line, sorted by the year of publication. The peak positions are represented with red symbol, if the authors attributed the dominant metastable state formed to the undistorted T phase, while the blue symbol is used to designate those reports where the authors attribute the dominant metastable phase to be T′ phase. Clearly, there is a very good consistency in the peak positions between different reports, independent of the claim concerning the phase type, T or T′. We have also marked by vertical lines ranges of frequencies calculated for the H phase (in black) and for the T′ phase (in blue). This makes it evident that all peak positions in chemically exfoliated samples correspond to either the stable H phase or the distorted T′ phase.

**Electronic Structure considerations:**

Electronic structures of different polymorphs of MoS$_2$ can be rationalized to a large extent in terms of their crystal structures based on the ligand field theory. As shown in Fig. 5a, the trigonal prismatic coordination of H phase MoS$_2$ splits the five *d*-orbitals of Mo into three groups of energies with the $d_z{}^2$ orbital as the lowest energy and separated from the remaining four orbitals by a large energy gap. The two *d*-electrons of Mo in MoS$_2$ occupies this lowest energy of $d_z{}^2$ orbital making it filled and separated by a large gap from the two sets of empty degenerate levels, namely, $d_{x^2-y^2}$, $d_{xy}$ and $d_{yz}$, $d_{zx}$. In contrast, the octahedral ligand field splitting in case of the undistorted T phase leads to the splitting of the five *d*-orbitals into a lower energy, triply degenerate $t_{2g}$ levels, separated by a large energy gap from the doubly degenerate $e_g$ levels. In this case, the two *d*-electrons of Mo, occupying the triply degenerate $t_{2g}$ levels will give rise to partially occupied states, implying a metallic nature of the system. However, the ground state of the $t_{2g}{}^3$ electronic configuration is triply degenerate and is, therefore, unstable towards Jahn-Teller distortions. Such distortions will split the triply degenerate $t_{2g}$ orbitals into a lower lying doubly degenerate orbital group and a higher energy singly degenerate orbital, as shown in third panel of Fig. 5a. One may then anticipate the formation of a semiconductor with a small band-gap, controlled by the Jahn-Teller distortions, in such a distorted phase, compared to the band-gap in the H phase determined by the large ligand field splitting. Depending on the specific distortions of the T phase, lifting the degeneracy of the ground state, one arrives at the various semiconducting T′, T″ and T‴ phases with small band-gaps. While this simplified MO diagram provides one with a qualitative expectation of the differing electronic structures for the various crystallographic polymorphs of $MoS_2$, the actual magnitudes of the band-gaps will be determined by the extensive dispersion of the local molecular orbitals in the periodic solid forming bands. This can even induce qualitative changes in the metallic/insulating properties if the dispersional widths of the relevant bands become comparable to or larger than the gaps in the MO energy level diagrams shown schematically in Fig. 5a. Band structures have been calculated by many groups for various polymorphic forms of $MoS_2$ using a variety of approximations based on different first principles approaches. Such calculations find that the H form of $MoS_2$ is indeed the lowest energy phase and T form constitutes an unstable phase, undergoing spontaneous distortions to T′, T″, T‴ structures, in broad agreement with the arguments presented above based on Fig. 5a.[12,38,39,69] The calculated band-gap of the H phase of $MoS_2$ shows both quantitative and qualitative dependency on the number of layers of $MoS_2$ involved with the band-gap varying from nearly 1.2 eV for the bulk to about 1.8 eV in the monolayer form; moreover, the nature of the band-gap is a direct one for the single layer, while multilayer H $MoS_2$ presents an indirect band-gap[82,83] with ample experimental validations[84,85] of these suggestions. Fig. 5b, showing the band dispersions for a single layer of H-$MoS_2$, exhibits the direct band-gap of 1.67 eV,[39] at the k points, consistent with other calculations.[13,86] The calculated band dispersions for a monolayer of the unstable T-$MoS_2$ show three bands crossing the Fermi level, confirming its metallic nature, as illustrated in Fig. 5c. The band structures of monolayers of distorted octahedral phases (T′, T″, T‴) of $MoS_2$ are shown in Fig. 5d - 5e. From these figures, it is evident that both T″ and T‴ phases have bandgaps (see Fig. 5e and 5f) of about 14 and 57 meV, significantly smaller than that in the H phase (Fig. 5b). The T′ appears to have a Dirac cone formed between B and Γ points, as illustrated in Fig. 5d. However, it has been shown by some authors that incorporating spin-orbit coupling (SOC) within the Mo 4$d$ in the calculation leads to the splitting of this Dirac cone and opening of a band-gap of ∼ 50 meV (Fig. 5g).[39] There are other reports suggesting the formation of similarly small band-gap (≤ 100 meV) for the T′ phase by several groups.[32,87] Since some of these calculations do not involve SOC, the essential role of SOC in forming the band-gap in the T′ phase is not fully established. SOC decreases the band-gap of the T″ structure, while the band-gap of the T‴ structure is almost unaffected by the SOC showing in Fig. 5h and 5i, respectively.

We now focus on experimental investigations leading to claims of formation of specific polymorphs, other than the stable H phase, due to chemical exfoliations of $MoS_2$ in the 2D forms. Apart from T and T′, mention of other metastable phases, such as T″ and T‴, is almost non-existent in the experimental literature. Therefore, the origin and the nature of unusual properties of chemically exfoliated 2D $MoS_2$ revolve around the primary issues: what is the most abundant additional phase formed, T or T′; and what is the electronic properties of that state? Is it metallic or semiconducting? Theoretical considerations, of course, suggest that T phase in the extended bulk form, is not even a metastable state but a dynamically unstable state;[38,40] in addition, the formation energy of the T phase is quite high compared to that of the T′ phase,[39] making the formation of T phase in preference over T′ phase unlikely. As already shown in Table 2, several claims of the formation of the T phase were based on additional Raman peaks appearing in the spectra of the exfoliated samples. Many of these publications interpret the T phase formation erroneously citing the original works in refs. 61 and 70, since these original publications already referred to the formation of (2×1) distorted T phase, which in today's terminology is the T′ phase. Since we have discussed the Raman results earlier in this article, we focus our attention to other experimental probes into the nature of the metastable states formed. While the erroneous conclusion of the T phase formation based on Raman studies led many groups to conclude, consequently, a metallic nature, there are also reports of some direct investigations of the electronic structure and transport properties that are relevant in the present context. For example, the results of current sensing atomic force microscopy (CSAFM) on $MoS_2$ before and after the chemical exfoliation are shown in Fig. 6a and 6b in terms of the representative conductivity maps.[42] Curiously, the conductivity map of the pristine H sample in Fig. 6a exhibits extreme inhomogeneity, with patches of highly insulating regions, marked black on a surprisingly conducting major phase on the surface of a well-known insulating H phase, possibly arising from the limitations of the technique or the sample. On the other hand, more relevant to the present discussion, the map of the chemically exfoliated sample in Fig. 6b presents a highly conducting homogeneous surface. I-V diagram, also shown in Fig. 6b, was used to assert the metallic nature of the chemically exfoliated sample. In passing, we note that I-V curve of the exfoliated sample in Fig. 6b is unusual with a highly non-ohmic, switching behaviour for an extremely small threshold voltage followed by current saturations, not expected of a usual metal. Despite these unusual aspects, these experiments point to a reasonable conductivity of the chemically exfoliated sample. Absorbance data of chemically exfoliated samples have also been used[33] to suggest the existence of the metallic T phase. This work[33] also reports the systematic change in the optical absorbance spectra, accompanying the conversion of the metastable phase to the stable one as a function of the annealing temperature. These absorbance spectra are shown in Fig. 6c. Spectral features, preferentially present in the sample with higher proportion of the large band-gap H phase following the higher temperature annealing, are interpreted as excitonic features. It was argued that the absence of any excitonic peak in the exfoliated sample at room temperature indicated that, the as synthesized, exfoliated sample was metallic and, consequently, existed in the T polymorphic form. However, as pointed out in Table 2, the as-synthesized sample in this case exhibited $J_1$ (151 cm$^{-1}$), $J_2$ (229 cm$^{-1}$), $E_{1g}$ (300 cm$^{-1}$) and $J_3$ (332 cm$^{-1}$) peaks in its Raman spectrum and this is inconsistent with high symmetry of the undistorted T phase. In this context, we note that there is a significant level of absorbance, extending to the longest wavelength displayed in Fig. 6c, for the chemically exfoliated sample without annealing. This featureless absorbance must be associated with the metastable phase formed, since the absorbance is found to decrease systematically with an increasing conversion of the metastable state to the stable H phase with a large band-gap (~ 1.8 eV or ~ 690 nm) with the successively higher annealing temperatures. This allows for the possibility of a small band-gap existing for the metastable phase, since the excitonic peak for that phase will appear in the vicinity of its band-gap. Considering that the various estimates of the band-gap of T′ phase is smaller than 100 meV, the spectral features in Fig. 6c do not exclude the possibility of the metastable state being T′ phase with its observable excitonic features lying outside of the wavelengths probed and presented in Fig. 6c.

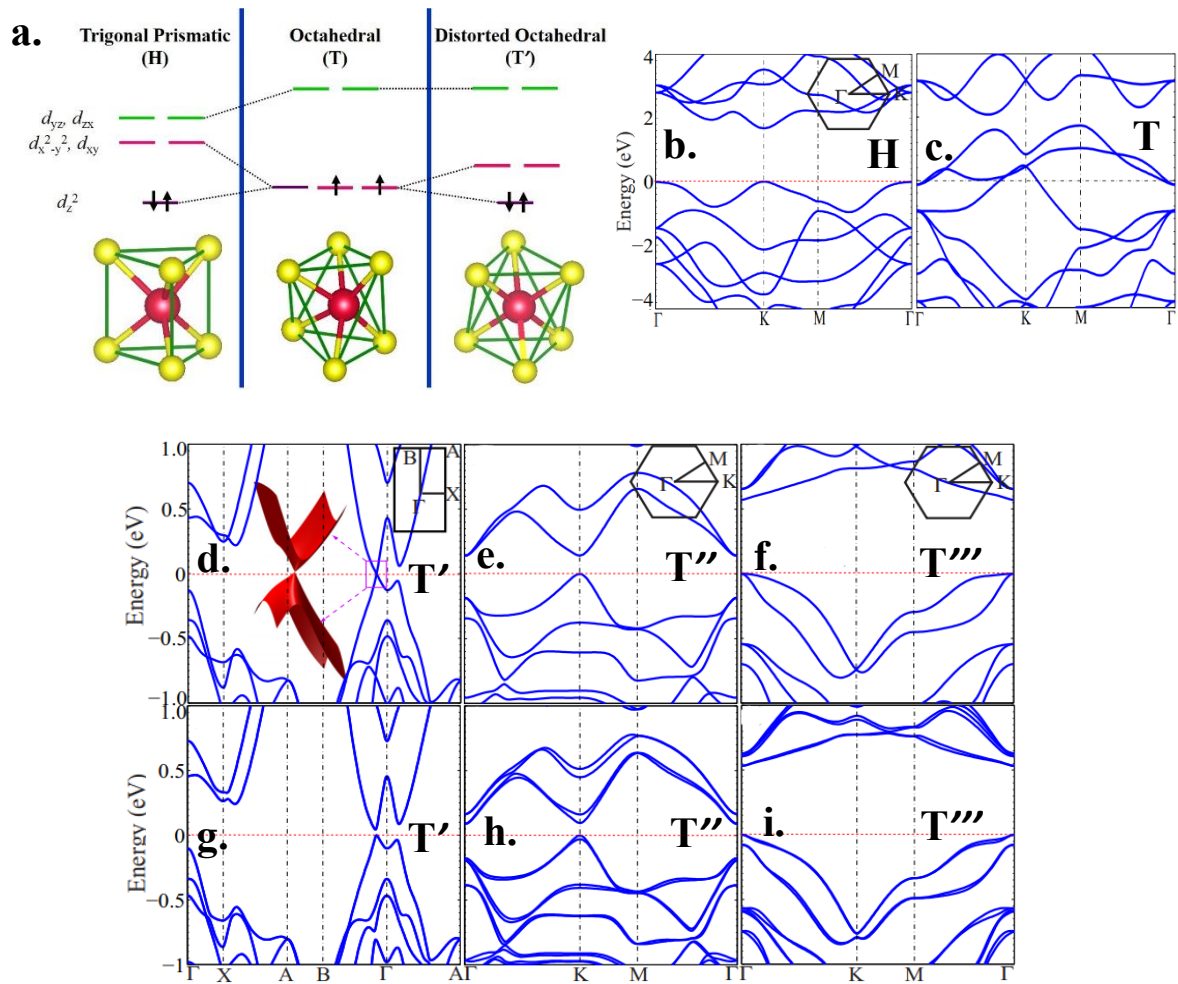

**FIG. 5.** (a) The building [$MoS_6$] unit of H, T′ and T phase with the corresponding $4d$ orbital splitting of central $Mo^{+4}$; Calculated electronic band structures of monolayer of (b) H phase, (c) T phase and different distorted octahedral phase i.e. (d) T′, (e) T″, (f) T‴. The corresponding band structures with SOC are shown in the lower panel for (g) T′, (h) T″, (i) T‴. Figures reproduced with permission from: (b) - (i), ref. 39, © 2017 APS.

It is well-known that the issue of metal/insulator property can be most easily probed by photoelectron spectroscopy, as it maps out electron states directly. Thus, it characterises a metal by the presence of a finite photoelectron spectral intensity at the Fermi energy, while an insulator is characterized by a finite energy gap between the Fermi energy and the onset of the finite spectral intensity. In order to enhance the contribution from the metastable phase to photoelectron spectra, we performed[32] scanning photoelectron microscopy experiment with a photon beam size down to 120 nm. The photon beam was positioned on the sample to maximize the contribution of the metastable state and the valence band spectra were obtained from the same spot. A magnified view of the energy region around the Fermi energy is shown in the inset of Fig. 6d. Clearly, the spectral intensity at the Fermi energy is negligibly small for both the H phases, termed meMoS$_2$, and the chemically exfoliated sample (ceMoS$_2$), establishing the semiconducting nature of the metastable state, consistent with the interpretation of the formation of the T′ phase. The above interpretation, however, does not explain why several past experiments found evidence of substantial conductivity for such metastable states. One possibility of course is that the thermally excited charge carriers are substantial for such a small band-gap semiconductor, whereas such charge carriers will be entirely negligible for the large band-gap H phase. In addition, we need to consider the possibility of charge carrier doping of such a small band-gap semiconductor as a plausible origin of the observed conductivity. In order to address this possibility, we prepared samples with different extent of Li$^+$ ions present in it by simply varying the water washing cycle after Li-intercalation.[78] The valence band spectra from the two extremes of washing, namely no washing at all (termed 0W) and after 12 cycles of washing with water (12W) are shown in Fig. 6e. These spectra with essentially zero intensity at the Fermi energy, establish both samples as small band gap semiconductors. We also find that the valence band spectrum obtained from 12W sample is shifted towards the higher binding energy side by almost 0.2 eV, indicating electron doping of MoS$_2$ by the Li ions. Therefore, it is possible that such charge doping of the chemically exfoliated MoS$_2$ samples contribute to the conduction though the polymorphic phase remains T′, as suggested by the Raman frequencies and low photoelectron spectral intensity at the E$_F$.

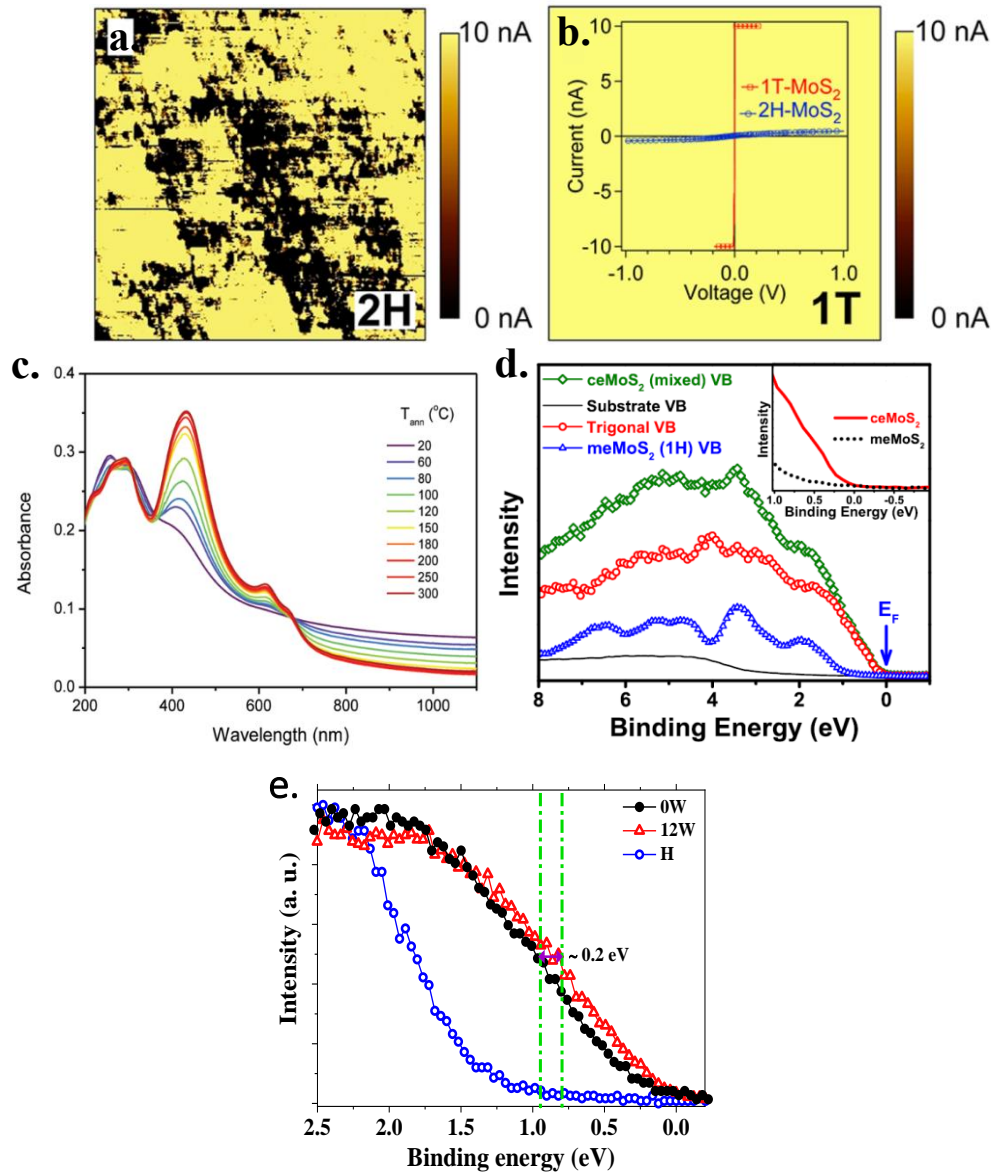

**FIG. 6.** Conductivity maps of **(a)** pristine MoS$_2$ and **(b)** chemically exfoliated MoS$_2$. Superimposed on the conductivity map in (b) is the I-V plot of corresponding samples; **(c)** Absorption spectra of lithium intercalated and exfoliated MoS$_2$ flakes at different annealing temperature; **(d)** Valence band spectra obtained from mechanically (meMoS$_2$) and chemically exfoliated (ceMoS$_2$) samples along with the calculated valence band of pure T' phase, inset showing the zoomed Fermi energy edge for both the samples; **(e)** Spectra near the Fermi edges for pristine H, and the samples with different extent of Li$^+$ present in it. Figures reproduced with permission from: **(a), (b),** ref. 42, © 2013 ACS; **(c)** ref. 33, © 2011 ACS; **(d)** ref. 32, © 2017 APS; **(e)** ref. 78, © 2020 Elsevier.

In summary, we first discussed different routes to chemical exfoliation of bulk MoS2 that provide the most convenient ways to synthesize copious amounts of 2D MoS2. Such chemical exfoliation has been shown to give rise to several polymorphs of MoS2 in addition to the most stable H phase and many interesting properties and device applications of such samples have been attributed in the past literature to the presence of these additional phases. Surveying the existing literature, we help to focus on the ambiguities present in identifying the dominant polymorphic phase present in such samples and show that the existing literature in terms of Raman spectra provides an overwhelming evidence in favour of the T' phase being present, rather than the often stated T phase. Since the T' phase is known to be semiconducting, we then address the puzzling issue of several more direct probes of the electronic structures of these samples appear to point to a highly conducting state of such exfoliated samples. We show that the substantial conductivity has to be understood in terms of thermal and dopant induced charge-carrier dopings of the small band-gap T' phase, rather than in terms of the formation of the unstable, metallic T phase.

**Acknowledgement:**

The authors thank SERB, Nanomission and DST, Government of India and Jamsetji Tata Trust for the support of this research. D.P. acknowledges the Council of Scientific and Industrial Research for a student fellowship.